# Does Geopolitics Have an Impact on Energy Trade? Empirical Research on Emerging Countries

Fen Li [1], Cunyi Yang [2,*], Zhenghui Li [3,*] and Pierre Failler [4]

[1] School of Marxism, Hunan Institute of Technology, Hengyang 421001, China; lifen@hnit.edu.cn
[2] School of Economics and Statistics, Guangzhou University, Guangzhou 510006, China
[3] Guangzhou Institute of International Finance, Guangzhou 510405, China
[4] Department of Economics and Finance, Portsmouth Business School, University of Portsmouth, Portsmouth P01 3DE, UK; pierre.failler@port.ac.uk
* Correspondence: 2112064077@e.gzhu.edu.cn (C.Y.); lizh@gzhu.edu.cn (Z.L.)

**Abstract:** The energy trade is an important pillar of each country's development, making up for the imbalance in the production and consumption of fossil fuels. Geopolitical risks affect the energy trade of various countries to a certain extent, but the causes of geopolitical risks are complex, and energy trade also involves many aspects, so the impact of geopolitics on energy trade is also complex. Based on the monthly data from 2000 to 2020 of 17 emerging economies, this paper employs the fixed-effect model and the regression-discontinuity (RD) model to verify the negative impact of geopolitics on energy trade first and then analyze the mechanism and heterogeneity of the impact. The following conclusions are drawn: First, geopolitics has a significant negative impact on the import and export of the energy trade, and the inhibition on the export is greater than that on the import. Second, the impact mechanism of geopolitics on the energy trade is reflected in the lagging effect and mediating effect on the imports and exports; that is, the negative impact of geopolitics on energy trade continued to be significant 10 months later. Coal and crude oil prices, as mediating variables, decreased to reduce the imports and exports, whereas natural gas prices showed an increase. Third, the impact of geopolitics on energy trade is heterogeneous in terms of national attribute characteristics and geo-event types.

## 1. Introduction

The energy trade is easily affected by geopolitical risks (GPR for short). The production and consumption of fossil fuels are unbalanced in various countries, which gives rise to the energy trade between countries. The complex international energy trade constitutes the existing complex international energy market. Since the 21st century, international political and economic events have occurred constantly and geopolitical risks have fluctuated for a long time, which have exerted a certain impact on the energy trade market and aroused the attention of scholars around the world. It is generally believed that political risk is a key factor to be considered in energy policy [1–3]. In recent years, emerging countries have gradually become the main players in the international energy market. At the same time, emerging countries are usually at the center of the geopolitical whirlpool. Although relevant studies have been quite rich, they usually only analyze the relationship between geopolitical risk and energy from the time dimension of years. At the same time, there are few specific studies on the impact mechanism and heterogeneity, and the conclusions are lacking accuracy and pertinence. Therefore, a more detailed study of the impact of geopolitical risks on the energy trade of emerging countries is necessary.

The main work and marginal contribution of this paper is to study the effect of geopolitical risk on energy trade, as well as mechanism research and heterogeneity analysis. The details are as follows: First, the impact of geopolitical risks on energy trade imports and exports as a whole is measured and verified. Through the fixed effect model with lag variables and the robustness test, excluding non-geopolitical factors, this paper finds that geopolitical risk has a certain inhibitory effect on energy imports and exports of emerging economies. Second, the impact mechanism of geopolitical risks on energy trade in emerging economies is studied. On the one hand, the negative impact on both energy exports and imports shows a short-term and medium-term lag; on the other hand, the three major energy prices play a mediating role in the geopolitical impact on the energy trade. Third, the heterogeneity of the geopolitical impact on the energy trade is studied. In a study of country attributes, on the one hand, OECD member countries' imports showed no significant positive performance in the rise of geopolitical risks, but their exports declined more; on the other hand, energy importing countries were more negatively affected in geopolitical disputes. In a heterogeneous study of geopolitics' impact on the energy trade of emerging economies based on the difference of geopolitical event types, this paper finds that there is indeed heterogeneity in geopolitical event types; that is, political events usually lead to an increase in the energy trade, whereas social events and economic events lead to a decrease in the energy trade.

The rest of this paper is structured as follows (see Figure 1 for the logical framework): The third part is the econometric test of the impact of geopolitical risk on the energy trade. The purpose is to empirically test the significance of the impact of geopolitical risk on the imports and exports of the energy trade through sample data, econometric models, and robustness tests. The fourth part is the mechanism analysis of the impact of geopolitical risk on the energy trade, considering the lag effect and the mediating effect. The fifth part is the heterogeneity analysis of the impact of geopolitical risk on the energy trade. The heterogeneity of the impact of geopolitical risk on the energy trade in different samples is studied first according to the national characteristics and attributes and then according to different types of geopolitical events. The sixth part draws the basic conclusions.

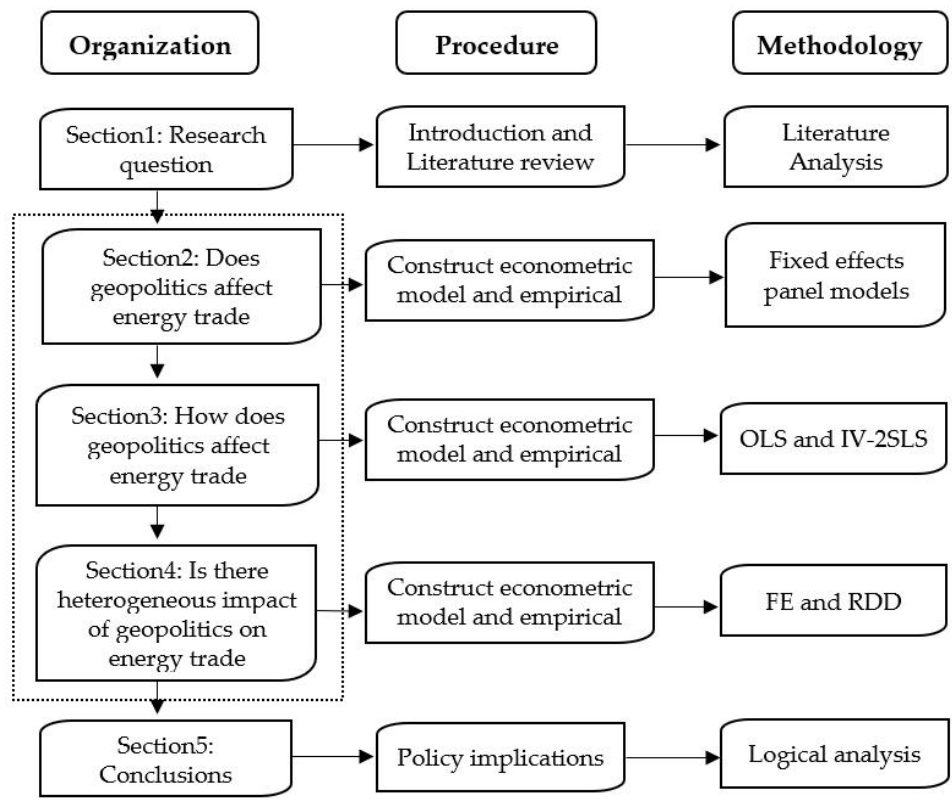

**Figure 1.** The logical framework of this paper.

## 2. Literature Review

Previous studies have shown that political risk affects energy trade in many ways. In terms of energy demand, the normal production activities of a country may be affected by the unstable macro environment, which will affect the energy input. Lee et al. used the SVAR framework and found that the country risk reduced the country's energy consumption, thus further reducing the country's dependence on global energy trade [2]; Kang and Ratti adopted the economic policy uncertainty index as a risk indicator and discovered that the positive impact on global real aggregate demand had a significant negative influence on the uncertainty of U.S. economic policy, whereas the impact on specific oil demand had an opposite effect [4]. Antonakakis et al. studied and showed that economic policy uncertainty (oil price shock) had a negative response to aggregate demand oil price shock (economic policy uncertainty shock) [5]. In terms of energy supply, high risks would lead to stagnation of national energy production and supply activities, thus failing to stably provide enough energy to meet domestic demand and foreign exports, such as in the Iran–Iraq War and the Ukraine crisis [6]. The supply side shock of global oil production did not have a significant impact on the economic policy uncertainty of the United States [4]. Zhang et al. found through gravity model analysis that importing countries were more willing to import liquefied natural gas (LNG) from regions with a stable political environment, which may because they want to ensure the security of their energy supply [7]. Oil importing countries are also concerned about energy supply security, as many oil exporting countries are characterized by high levels of political instability [8]. In terms of energy transactions, political risks can also affect global energy trade by influencing the energy transport process, energy investment, energy prices, and other related factors [2,3,9–12].

Frontier and emerging countries tend to play different roles when geopolitical risks occur. In the analysis of frontier countries, Mercille and Jones, who outlined radical geopolitics, believe that America follows the geopolitical logic of "resolutely resisting any challenge to American hegemony" to prevent the situation of "falling dominos," "apples in a barrel infected by one rotten one," or "a growing cancer," and the EU has abandoned the agreement on the Iranian nuclear issue and failed to fulfill its obligations [13]. Some geographers also discuss the role of oil in the US intervention in the Middle East [14,15]. The geopolitical and geo-economic group (the Three Seas Initiative), mainly composed of post-communist NATO and EU member states, is highly dependent on NATO as a protector, which is actually the U.S. protective umbrella [16]. The outbreak of the European economic crisis foresaw the recovery of geopolitics in the analysis of trans-Atlantic foreign policy, and the European Union was still in danger [17]. The development of political framework conditions is currently facing far-reaching challenges in Europe [18]. The profit is disproportionately distributed to the actors at higher levels in the supply chain [19]. Some structural problems have increased prominently in emerging countries in recent years. In the face of geopolitical risks, emerging countries also face disagreement. Ferdinand believes that from the perspective of geopolitics, although China continues to maintain a high degree of domestic political stability, it has changed from a risk-averse country to a risk-accepting country abroad [20]; China and Russia are to some extent against sanctions imposed by Western countries [13]. According to information geopolitics, Russia does not regard Internet governance, cyber security, or media policy as separate fields; on the contrary, all fields covered by these disciplines fall under the category of "information security" in Russia's foreign policy and are strategically used to achieve geopolitical goals [21]. Oral and Ozdemir believe that the political and economic policies carried out by big powers for the sake of energy security put Turkey in a predicament [22]. As a link in the political triangle between China, the United States, and India, India can play its own balancing game to maximize its own interests. India's pursuit of global power status will be promoted through its efforts to achieve economic growth and military goals [23]. Suk believes that South Korea is in a geopolitical environment and is forced by neighboring powers to make a zero-sum choice, which actually means that South Korea's foreign choice is becoming an important factor determining the power map of the Eurasian continent [24].

When geopolitical risks occur, different types of countries will show different reactions in the energy trade according to the degree of their own interest correlation. In their research on frontier countries and emerging countries, Simonia and Torkunov stated that the main factor affecting the pricing of the global energy industry is geopolitics, and the main "volcano" of its turbulence is the United States [25]. The member states of the Three Seas

Initiative are determined to reduce their dependence on natural gas from Russia and Ukraine [16]. Oral and Ozdemir believe that since 70% of the world's oil and gas reserves are located near Turkey, in the context of energy geopolitics, Turkey's most important goal is to become the global energy trade center [22]. Oswald discussed the way out when Mexico faces political risks and held that the oil price crisis opens the possibility for Mexico to promote its abundant renewable energy potential [26]. When peak oil or geopolitical issues drive oil prices to unrealistic levels, the industrialization of shale oil will eventually occur, and for Brazil, shale oil can be used as a strategic resource [27]. Another important classification of energy trade is from the aspect of energy importing and exporting countries. In a study on the heterogeneity of energy importing and exporting countries, Lee et al. found that the unexpected positive impact of oil prices reduced the country risk of net oil-exporting countries and increased the country risk of net oil-importing countries [2]. The export volume is mainly affected by oil reserves and domestic exports, whereas the import volume is mainly affected by the economic growth and oil consumption of importing countries [28]. The country risk has a great impact on the trade patterns of energy-importing and -exporting countries, and importing countries should pay attention to the negative impact of economic risks, which reduces the importers' anti-control ability on resources and worsens its relationship with important countries. For exporting countries, political and economic risks have a great negative impact on their total trade volume and resource control ability [29].

The above studies provided reference experience and new ideas for this research.

**3. Econometric Examination of the Impact of Geopolitics on the Energy Trade**

*3.1. Model Specification*

The occurrence of geopolitical events will lead to the rise of geopolitical risks. The severity of geopolitical risks will affect the normal activities of emerging economies and inhibit energy trade among countries by impeding energy production, consumption, and transportation. At the same time, the production and consumption of energy are unbalanced in each country, which leads to the difference between energy imports and exports. If the global energy trade is simply regarded as a "zero-sum trade" without external impetus, emerging economies, frontier economies, and backward economies will reach the equilibrium point of their energy imports and exports due to their development stages and their positions in the energy market. In the current world of close connection, when local geopolitical risks appear, countries in the risk vortex will be directly impacted, and countries with a low correlation with the event will also show specific performance in energy import and export due to political turmoil or energy price fluctuations in associated countries. In general, instability will lead to the inhibition of production related to energy input, and the energy production and supply chain will also be hindered.

In the past, international trade studies usually considered trade cost factors, such as distance, transportation cost, and tax [30–32], but this study only examined the geopolitical risk and energy import and export trade volume of an economy as a whole, and did not distinguish the risk and trade between an economy and a single object. If we can measure the geopolitical risk between the target economy and the specific object, this may be the direction of further research in the future.

This paper adopts a fixed effect regression model with lag explanatory variables to test the impact of the geopolitical risk index on energy imports and exports of emerging economies. Compared with the OLS estimation, the fixed effect model of time and individual can simultaneously control the influence of regional fixed factors that do not change with time and macro factors that do not change with region on the regression results. The regression model is as follows:

$$LnImport_{it} = \lambda_0 + \lambda_1 * LnGPR_{it} + \lambda_2 * LnGPR_{i(t-1)} + \lambda_3 * LnGPR_{i(t-2)} + \lambda_4 * LnGPR_{i(t-3)} + \lambda_5 X_{it} + \eta_t + \pi_i + \varepsilon_{it} \quad (1)$$

$$LnExport_{it} = \beta_0 + \beta_1 * LnGPR_{it} + \beta_2 * LnGPR_{i(t-1)} + \beta_3 * LnGPR_{i(t-2)} + \beta_4 * LnGPR_{i(t-3)} + \beta_5 X_{it} + \eta_t + \pi_i + \varepsilon_{it} \quad (2)$$

In Formulas (1) and (2), i represents individual country and t is the time of the month; LnImport and LnExport are the explained variables, which are the logarithm of energy imports and energy exports, respectively; and LnGPR is the explanatory variable. Considering the lag effect of the impact, the LnGPR of (t−1)/(t−2)/(t−3) is included in the regression; X is the covariate that may affect the imports and exports, including the current GDP, real interest rate, and USD exchange rate of individual i at time t. LnGDP is the covariate and the logarithm of

| Sample | Item | GPR | LnImport | LnExport | LnGDP |
|---|---|---|---|---|---|
| Summary | N | 4284 | 4009 | 4009 | 4251 |
| | Mean | 99.38 | 20.58 | 20.74 | 10.70 |
| | Std | 34.66 | 1.82 | 1.87 | 1.03 |
| Argentina | N | 252 | 240 | 240 | 252 |
| | Mean | 97.72 | 19.24 | 19.64 | 10.30 |
| | Std | 34.89 | 1.03 | 0.78 | 0.51 |
| Brazil | N | 252 | 240 | 240 | 252 |
| | Mean | 104.52 | 21.22 | 20.72 | 11.65 |
| | Std | 30.31 | 0.61 | 0.84 | 0.55 |
| Russia | N | 252 | 240 | 240 | 252 |
| | Mean | 107.79 | 19.01 | 23.37 | 11.44 |
| | Std | 29.21 | 0.50 | 0.54 | 0.64 |
| Philippines | N | 252 | 240 | 240 | 252 |
| | Mean | 104.70 | 20.38 | 18.04 | 9.62 |
| | Std | 34.30 | 0.49 | 0.66 | 0.55 |
| Colombia | N | 252 | 240 | 240 | 252 |
| | Mean | 81.52 | 18.65 | 20.92 | 9.80 |
| | Std | 31.28 | 1.25 | 0.71 | 0.49 |
| South Korea | N | 252 | 240 | 240 | 252 |
| | Mean | 110.53 | 22.78 | 21.47 | 11.39 |
| | Std | 43.14 | 0.54 | 0.70 | 0.36 |
| Malaysia | N | 252 | 240 | 240 | 252 |
| | Mean | 94.16 | 20.93 | 21.48 | 9.81 |
| | Std | 36.68 | 0.70 | 0.55 | 0.45 |
| Mexico | N | 252 | 240 | 240 | 252 |
| | Mean | 110.71 | 21.21 | 21.60 | 11.33 |
| | Std | 26.04 | 0.73 | 0.46 | 0.21 |
| South Africa | N | 252 | 240 | 240 | 252 |
| | Mean | 89.10 | 20.68 | 20.10 | 10.03 |
| | Std | 29.53 | 0.61 | 0.44 | 0.38 |
| Saudi Arabia | N | 252 | 216 | 216 | 234 |
| | Mean | 102.90 | 17.22 | 23.38 | 10.58 |
| | Std | 29.24 | 1.34 | 0.54 | 0.49 |
| Thailand | N | 252 | 240 | 240 | 252 |
| | Mean | 95.96 | 21.48 | 20.13 | 10.08 |
| | Std | 42.01 | 0.60 | 0.65 | 0.48 |
| Turkey | N | 252 | 240 | 240 | 252 |
| | Mean | 118.11 | 21.62 | 19.45 | 10.81 |
| | Std | 39.61 | 0.60 | 0.83 | 0.48 |
| Venezuela | N | 252 | 228 | 228 | 240 |
| | Mean | 104.09 | 17.74 | 21.87 | 9.83 |
| | Std | 34.71 | 1.11 | 0.72 | 0.50 |
| Israel | N | 252 | 228 | 228 | 252 |

|  | | Mean | 90.77 | 20.29 | 16.25 | 9.82 |
|---|---|---|---|---|---|---|
|  | | Std | 22.06 | 0.50 | 1.91 | 0.41 |
| India | | N | 252 | 228 | 228 | 252 |
|  | | Mean | 91.23 | 22.61 | 21.27 | 11.58 |
|  | | Std | 28.00 | 0.80 | 1.08 | 0.60 |
| Indonesia | | N | 252 | 229 | 229 | 252 |
|  | | Mean | 77.61 | 21.27 | 21.70 | 10.74 |
|  | | Std | 31.52 | 0.63 | 0.48 | 0.63 |
| China | | N | 252 | 240 | 240 | 252 |
|  | | Mean | 107.96 | 23.11 | 21.38 | 12.95 |
|  | | Std | 30.67 | 0.95 | 0.52 | 0.86 |

GDP. In this paper, the monthly GDP of each country is approved by frequency conversion of its quarterly GDP (source: World Bank), without changing the trend. $\eta_t$ is the time fixed effect, $\pi_i$ is the individual fixed effect, and $\varepsilon_{it}$ is the error term.

*3.2. Variables and Data*

Seventeen emerging economies were selected as the initial research samples. For the purpose of more detailed verification of the impact, as well as for the availability of the data, this paper includes monthly data from January 2000 to December 2020, most of which actually start from January 2001 due to the lack of data. The GPR index, put forward by Caldara and Iacoviello [33], quantifies the geopolitical risks of 19 emerging economies through the analysis of newspaper articles from a specific period of time. It now provides monthly data from January 1985 to February 2021, and it also provides two decomposition indicators GPT and GPA, which will not be discussed in this paper since they only focus on global data. Considering the availability of data (the monthly data of the energy trade between Ukraine and Hong Kong are conspicuously missing), this paper selected 17 samples: Turkey, Mexico, South Korea, Russia, India, Brazil, China, Indonesia, Saudi Arabia, South Africa, Argentina, Colombia, Venezuela, Thailand, Israel, Malaysia, and the Philippines. Data on energy imports and exports (USD) came from the 27th category of the commodity codes HS from the International Trade Centre. This category includes coal, coke and coal bricks, petroleum, petroleum products and related raw materials, natural gas, and man-made gas and electric current, which can represent the overall energy import and export level of the economy. A country's monthly GDP (million USD) is calculated by the frequency conversion of its quarterly GDP (source: World Bank) without changing its trend. Current interest rates and exchange rates of sample countries were obtained from Wind and EPU databases. The logarithm of trade volume is used in most studies to mitigate the fluctuation trend of data and to alleviate heteroscedasticity to a certain extent, because the trade volume gap between countries is generally large and the discrete trend of data is strong. According to the standard practice, this paper adopted a logarithmic treatment for the above data (except exchange rate and interest rate). Data processing was completed by SPSS 24.0, Stata16, and Eviews11 software. Table 1 reports the descriptive statistical results of the variables.

**Table 1.** Descriptive statistics.

*3.3. Results of the Econometric Tests*

3.3.1. Empirical Results

This paper analyzed the impact of geopolitical risks on energy imports and exports of emerging economies by using the fixed effect regression model (1) and (2) with lag variables, especially focusing on the coefficient size and significance of the explanatory variable LnGPR. In addition, a more stringent fixed effect PPML test was conducted to verify the empirical results. The parameter estimation results from the model are shown in Table 2.

**Table 2.** The impact of GPR on energy imports and exports.

|  | LnImport | | | LnExport | | |
|---|---|---|---|---|---|---|
|  | (1) | (2) | (3) | (4) | (5) | (6) |
| LnGPR | −0.146 * (−1.70) | −0.069 ** (−2.19) | −0.004 *** (−2.57) | −0.036 (−0.37) | −0.092 ** (−2.52) | −0.004 *** (−2.71) |
| LnGPR _lag1 | −0.085 (−0.92) | −0.019 (−0.60) | −0.001 (−0.74) | −0.015 (−0.14) | −0.048 (−1.23) | −0.002 (−1.40) |
| LnGPR_lag2 | −0.081 (−0.88) | −0.024 (−0.74) | −0.001 (−0.87) | −0.029 (−0.28) | −0.060 (−1.54) | −0.003 * (−1.73) |
| LnGPR_lag3 | −0.129 (−1.50) | −0.027 (−0.91) | −0.002 (−1.09) | −0.070 (−0.71) | −0.109 *** (−2.99) | −0.005 *** (−3.28) |
| LnGDP | 1.056 | 1.272 | 0.062 | 0.821 | 0.949 | 0.045 |
| Interest rate | −0.016 | 0.002 | −0.000 | −0.020 | −0.018 | −0.000 |
| FX rate | 0.000 | −0.000 | 0.000 | 0.000 | −0.000 | −0.001 |
| Cons | 11.332 | 7.693 | 2.353 | 12.671 | 12.226 | 2.594 |
| Time control | No | Yes | Yes | No | Yes | Yes |
| Individual control | No | Yes | Yes | No | Yes | Yes |
| N | 4009 | 4009 | 4009 | 4009 | 4009 | 4009 |
| F | 353.22 | 934.54 | - | 184.81 | 451.31 | - |
| R-squared | 0.3819 | 0.3184 | 0.0299(Pseudo) | 0.2443 | 0.1438 | 0.0309(Pseudo) |

[1] Notes: *, **, *** stand for significant levels of 10%, 5%, and 1%, respectively, and the values in brackets are T-values.

As can be seen from Table 2, the GPR index had a significant negative impact on the energy imports and exports of emerging economies. First of all, the rising geopolitical risks could suppress emerging energy imports and exports, indicating that when geopolitical risk occurs, the energy trade of emerging economies may affect the demand for energy input due to the decline of domestic production [2], resulting in the stagnation of national energy production and supplies so enough energy cannot be reliably provided to meet foreign export supply [24]. Or it can inhibit global energy trade by affecting the energy transport process, energy investment, energy prices, and other related factors [2,3,9]. Secondly, the negative impact of geopolitical risk on energy exports of emerging economies was greater than that on energy imports. If the global energy trade is simply treated as a "zero-sum trade," it can account for the negative impact on energy imports being greater than that on exports when countries outside emerging economies are under the influence of geopolitical risks. Thirdly, the restraining effect of geopolitical risk on energy trade volume was not only reflected in the immediate occurrence of geopolitical risk, but also had a certain degree of lag effect in the short term. The FE-PPML Models (3) and (6) verify the above conclusions.

3.3.2. Robustness Tests

First, we'll address the test on non-geopolitical factors. The outbreak of the financial crisis in 2008 affected the entire financial system and even the entire economic system, regardless of geographical location or market development level [34]. Conceptually, the crisis was not a geopolitical event, but after its outbreak, global energy prices plummeted, and the energy trade market also suffered a certain impact. In order to exclude the possibility that the change in energy trade volume was affected by the financial crisis, this paper took the outbreak of the financial crisis as the dummy variable Et; that is, before September 2008, Et=0, and after September 2008, Et=1. The model is as follows:

$$LnImport_{it}/LnExport_{it} = \lambda_0 + \lambda_1 * LnGPR_{it} + \lambda_2 * E_t + \lambda_3 * X_{it} + \eta_t + \pi_i + \varepsilon_{it} \qquad (3)$$

The variables in Formula (3) are consistent with the above, and the regression results are shown in Table 3.

Table 3. The impact of GPR on energy imports and exports with Et.

|  | LnImport | LnExport |
|---|---|---|
| LnGPR | −0.128 *** (−5.16) | −0.184 *** (−6.06) |
| Dummy variable | Yes | Yes |
| Control variable | Yes | Yes |
| Time control | Yes | Yes |
| Individual control | Yes | Yes |

[1] Notes: *, **, *** stand for significant levels of 10%, 5%, and 1%, respectively, and the values in brackets are T-values.

As can be seen from Table 3, the direction and significance of the impact of geopolitical risk on energy trade volume are consistent with the previous conclusion, which is robust.

Second, we will address the bootstrap test. The bootstrap method is used for resampling with replacement to get more progressive and effective estimators. In this paper, 2000 repeated samples were used for regression coefficients, and the regression results are shown in Table 4.

**Table 4.** The impact of GPR on energy imports and exports with bootstrap.

|  | **LnImport** | **LnExport** |
|---|---|---|
| LnGPR | −0.103 *** (−4.20) | −0.176 *** (−6.39) |
| Control variable | Yes | Yes |
| Time control | Yes | Yes |
| Individual control | Yes | Yes |

[1] Notes: *, **, *** stand for significant levels of 10%, 5%, and 1%, respectively, and the values in brackets are Z-values.

As can be seen from Table 4, the direction and significance of the impact of geopolitical risk on energy trade volume are consistent with the previous conclusion, which is robust.

Third, we will address the grouping regression test. In order to eliminate the possibility that periods of high and low geopolitical risks may differ in their impact on energy trade, this paper conducted a group regression on GPR index referring to the idea of quantile regression, considering the GPR indices of 0–10%, 10–25%, 25–50%, 50–75%, 75–90%, and 90–100%. The regression results are shown in Table 5.

**Table 5.** The impact of GPR on energy imports and exports with regression by group.

| Quantile of GPR | 0–10% | 10–25% | 25–50% | 50–75% | 75–90% | 90–100% |
|---|---|---|---|---|---|---|
|  | LnImport | | | | | |
| LnGPR | 0.048 | −0.332 | 0.060 | −0.154 | −0.926** | −0.045 |
|  | LnExport | | | | | |
| LnGPR | 0.083 | 0.241 | −0.607* | −0.587* | −0.528 | 0.082 |
| N | 400 | 603 | 1005 | 1000 | 600 | 401 |
| Control variable | Yes | | | | | |
| Time control | Yes | | | | | |
| Individual control | Yes | | | | | |

[1] Notes: *, **, *** stand for significant levels of 10%, 5%, and 1%, respectively.

As can be seen from Table 5, the overall direction of the impact of geopolitical risk on the energy trade is basically consistent with the previous conclusion, and the conclusion is still robust. In addition, in the period of middle and high geopolitical risk, geopolitical risk had the greatest negative impact on energy import. In the period of moderate geopolitical risk, geopolitical risk had the greatest negative impact on energy export.

**4. Mechanism Study on the Impact of Geopolitics on the Energy Trade**

*4.1. Time Lag Analysis*

Energy trade involves many factors, such as the international environment, geographical location, national strategy, and resource distribution of each country. Therefore, the mechanism through which geopolitical risks affect the energy trade of emerging economies is different. From the perspective of time lag, emerging economies usually maintain a high growth rate, their demand for energy input is large, and their import orders are rigid, which are difficult to shrink in the short term, whereas other countries (such as frontier countries) respond more quickly to energy trade, so the inhibition effect on the import volume of emerging economies is weak in the short term, and the short-term and medium-term impact shows an upward trend, but the total impact is small compared to that on the export volume. This paper conducted a 1–15-month lag test on energy imports and exports of emerging economies. The estimated results are shown in Table 6.

Table 6. The time lag effect of GPR on energy imports and exports.

|  | LnImport | LnExport |
|---|---|---|
|  | (1) | (2) |
| LnGPR _lag1 | −0.051 ** (−2.06) | −0.124 *** (−4.00) |
| LnGPR _lag2 | −0.048 * (−1.91) | −0.127 *** (−4.11) |
| LnGPR _lag3 | −0.044 * (−1.77) | −0.139 *** (−4.47) |
| LnGPR _lag4 | −0.047 * (−1.88) | −0.137 *** (−4.44) |
| LnGPR _lag5 | −0.054 ** (−2.15) | −0.140 *** (−4.53) |
| LnGPR _lag6 | −0.065 *** (−2.62) | −0.125 *** (−4.03) |
| LnGPR _lag7 | −0.066 *** (−2.64) | −0.102 *** (−3.31) |
| LnGPR _lag8 | −0.096 *** (−3.85) | −0.107 *** (−3.47) |
| LnGPR _lag9 | −0.104 *** (−4.17) | −0.106 *** (−3.44) |
| LnGPR _lag10 | −0.066 *** (−2.67) | −0.099 *** (−3.21) |
| LnGPR _lag11 | −0.069 *** (−2.77) | −0.100 *** (−3.23) |
| LnGPR _lag12 | −0.055 ** (−2.19) | −0.076 ** (−2.44) |
| LnGPR _lag13 | −0.038(−1.50) | −0.060 ** (−2.28) |
| LnGPR _lag14 | −0.017 (−0.69) | −0.023 (−1.17) |
| LnGPR _lag15 | −0.007 (−0.27) | −0.011 (−0.58) |
| Control variable | Yes | Yes |
| Time control | Yes | Yes |
| Individual control | Yes | Yes |

[1] Notes: *, **, *** stand for significant levels of 10%, 5%, and 1%, respectively, and the values in brackets are T-values.

It can be seen from Table 6 that the inhibiting effect of GPR on the energy imports and exports of emerging economies had a significant time-lag effect in the short and medium term, but the impact on the imports was weak in the short term, while showing an upward trend in the short and medium term. The impact was smaller than that on export. Through regression with more lag periods, it was found that the inhibiting effect of GPR on energy imports and exports of emerging economies disappeared in about 12–13 months. It should be noted that the above results were obtained by regression for each lag period. When regression was used as explanatory variable at the same time, although the regression was significantly decreased, it still showed a long-time lag effect.

*4.2. Mediating Effect Analysis*

From the perspective of energy price, previous studies usually used fossil fuels to represent energy, including coal, crude oil, and natural gas with different forms and quality levels [35,36]. Three fossil energy prices affect the energy supply and demand and turnover. Coal and crude oil were the earliest energy sources to be exploited and used, and play a crucial role in industry and transportation. However, natural gas is difficult to exploit, transport, and store; its calorific value is low [37]; and it is weaker than the former two in industry, so its mediating role may be different from the former two.

A mediating effect model [38] was constructed. Whether from the perspective of logic or economics, there are endogenous problems between geopolitical risk and energy price. Therefore, this paper combined the IV-2SLS method, taking LnGPR_lag1 as the tool variable of the endogenous explanatory variable LnGPR to verify the impact of coal price on energy imports and exports in emerging economies. In this paper, the monthly FOB of the steam coal spot in Newcastle and Kembla Port, Australia, was selected to represent the overall coal price, which is expressed as Pc (price of coal). Taking the study of import volume as an example, the mediating effect model is as follows:

$$LnImport_{it} = \lambda_0 + \lambda_1 * LnGPR_{it} + \lambda_2 * X_{it} + \varepsilon_{it} \quad (4)$$

$$LnP_{ct} = \alpha_0 + \alpha_1 * LnGPR_{it} + \alpha_2 * X_{it} + \varepsilon_{it} \quad (5)$$

$$LnImport_{it} = \gamma_0 + \gamma_1 * LnGPR_{it} + \gamma_2 * LnP_{ct} + \gamma_3 * X_{it} + \varepsilon_{it} \quad (6)$$

where $\lambda 1$ measures the impact of geopolitical risk on the import volume of emerging economies, $\alpha 1$ measures the impact of geopolitical risk on coal price Pc, $\gamma 1$ measures the direct impact of geopolitical risk on the imports of emerging economies, and $\gamma 2$ measures the impact of coal price Pc on the imports of emerging economies. In this paper, the regression coefficient was tested step by step. The meaning and calculation method of variables in the regression equations were consistent with those in the discontinuity regression. Due to the limitation of space, the mediating effect model with LnExport as the explanatory variable and Pco and Png as the mediating variables will not be repeated, and the form is consistent with Formulas (4)–(6). The regression results are shown in Table 7.

**Table 7.** Test results of the impact mechanism of GPR on energy imports and exports (Pc).

|                  | LnPc (1)    | LnImport (2) | LnImport (3) | LnExport (4) | LnExport (5) |
|------------------|-------------|--------------|--------------|--------------|--------------|
| LnGPR            | −0.423 ***  | −0.617 ***   | −0.455 ***   | −0.494 ***   | −0.279 **    |
| LnPc             | —           | —            | 0.341 ***    | —            | 0.453 ***    |
| Control variable | Yes         | Yes          | Yes          | Yes          | Yes          |
| N                | 4009        | 4009         | 4009         | 4009         | 4009         |
| R-squared        | 0.1618      | 0.3691       | 0.3777       | 0.2158       | 0.2285       |

[1] Notes: *, **, *** stand for significant levels of 10%, 5%, and 1%, respectively.

As can be seen from Table 7, Model (2) verified the direct impact of geopolitical risk on the imports of emerging economies. It can be seen from the results that the impact of geopolitical risk on the imports was significantly negative, which is consistent with the benchmark regression results. Model (1) verified the impact of geopolitical risk on coal price Pc, and the regression coefficient was −0.423 and significantly negative, indicating that geopolitical risk leads to a reduction in Pc. In Model (3), the coefficient of the impact of geopolitical risk on the imports of emerging economies was −0.455, which was significant. The impact of coal price Pc was significantly positive on the imports of emerging economies at a confidence level of 1%, which means that a rise in coal prices will lead to a rise in energy imports. In conclusion, it shows that the geopolitical risk suppressed the imports of emerging economies partly through the reduction of coal prices. Similarly, combining Models (1), (4), and (5), it is indicated that the suppression of geopolitical risk on the exports of emerging economies was partly realized by reducing coal prices. At the same time, this test showed that the supply and demand of coal markets in emerging economies are inelastic [39].

Then, the Sobel test was used to verify the robustness of the mediating effect, and the analysis results are consistent with the stepwise test. The total effect of GPR index on the energy imports of emerging economies was −0.388, equal to a direct effect of −0.286 plus a indirect effect of −0.102. The calculated mediating effect accounted for 26.31% of the total effect, which means that coal price played a partial mediating role in the influence of the GPR index on the energy imports of emerging economies. The total effect of the GPR index on the energy exports of emerging economies was −0.306, equal to a direct effect of −0.177 plus an indirect effect of −0.130. The calculated mediating effect accounted for 42.34% of the total effect, which means that the coal price played a strong mediating role in the impact of the GPR index on the energy exports of emerging economies.

A mediating effect model was constructed to verify the impact of crude oil price on energy imports and exports of emerging economies. Li et al. found that there was a strong correlation between geopolitical factors and crude oil prices during a period of political tension [40]. Faria et al. proved the positive correlation between China's export growth and oil prices by using a theoretical model [41]. In this paper, the monthly FOB of the Brent crude oil DTD spot at the main port of the United Kingdom was chosen to represent the overall crude oil price, which is expressed as Pco (price of crude oil). The regression results are shown in Table 8.

Table 8. Test results of the impact mechanism of GPR on energy imports and exports (Pco).

|  | LnPco (1) | LnImport (2) | LnImport (3) | LnExport (4) | LnExport (5) |
| --- | --- | --- | --- | --- | --- |
| LnGPR | −0.425 *** | −0.617 *** | −0.429 *** | −0.494 *** | −0.220 * |
| LnPco | — | — | 0.393 *** | — | 0.572 *** |
| Control Variable | Yes | Yes | Yes | Yes | Yes |
| N | 4009 | 4009 | 4009 | 4009 | 4009 |
| R-squared | 0.1320 | 0.3691 | 0.3809 | 0.2158 | 0.2368 |

[1] Notes: *, **, *** stand for significant levels of 10%, 5% and 1% respectively.

Table 8, combining Models (1), (2), and (3), shows that the suppression of geopolitical risk on the imports of emerging economies was partly achieved by lowering the price of crude oil. Similarly, combining Models (1), (4), and (5), it is shown that the suppression of geopolitical risk on the exports of emerging economies was partly realized by lowering the price of crude oil. At the same time, this test illustrated the inelasticity of supply and demand in the crude oil market of emerging economies [42].

The Sobel test was used again to verify the robustness of the mediating effect, and the analysis results are also consistent with the stepwise test. The total effect of GPR index on the energy imports of emerging economies was −0.388, equal to a direct effect of −0.272 plus an indirect effect of −0.116. The calculated mediating effect accounted for 29.89% of the total effect, which means that the crude oil price played a partial mediating role in the impact. The total effect of GPR index on the exports of emerging economies was −0.306, equal to a direct effect of −0.144 plus an indirect effect of −0.162. The calculated mediating effect accounted for 52.88% of the total effect, which means that the crude oil price played a strong mediating role in the impact of GPR index on the energy exports of emerging economies.

In order to verify the impact of natural gas price on energy imports and exports of emerging economies, a mediating effect model was constructed. In this paper, the monthly FOB price of the Russian produced natural gas spot from German ports was selected to represent the overall natural gas price, expressed as Png (price of natural gas). The regression results are shown in Table 9.

Table 9. Test results of the impact mechanism of GPR on energy imports and exports (Png).

|  | LnPng (1) | LnImport (2) | LnImport (3) | LnExport (4) | LnExport (5) |
| --- | --- | --- | --- | --- | --- |
| LnGPR | 0.104 ** | −0.617 *** | −0.524 ** | −0.494 *** | −0.321 *** |
| LnPng | - | - | −0.120 *** | - | −0.066 *** |
| Control variable | Yes | Yes | Yes | Yes | Yes |
| N | 4009 | 4009 | 4009 | 4009 | 4009 |
| R-squared | 0.0295 | 0.3691 | 0.3936 | 0.2158 | 0.2173 |

[1] Notes: *, **, *** stand for significant levels of 10%, 5%, and 1%, respectively.

By observing Table 9 and combining with Models (1), (2), and (3), it is shown that the suppression of geopolitical risk on the energy imports of emerging economies was partly realized through the increase of natural gas prices. Similarly, in combination with Models (1), (4), and (5), it is shown that the suppression of geopolitical risk on the energy exports of emerging economies was partly realized through the increase of natural gas prices. At the same time, this test showed that the supply and demand of natural gas markets in emerging economies are elastic [42,43].

Then, the Sobel test was adopted to verify the robustness of the mediating effect, and the analysis results are consistent with the stepwise test. The total effect of GPR index on the energy imports of emerging economies was −0.388, equal to a direct effect of −0.212 plus an indirect effect of −0.176. The calculated mediating effect accounted for 45.36% of the total effect. This means that natural gas price played a strong mediating role in the impact of the GPR index on the amount of energy imports of emerging economies. The total effect of GPR index on the energy exports of emerging economies was −0.306, equal to a direct effect of −0.257 plus an indirect effect of −0.049. The calculated mediating effect accounted for 16.01% of the total effect. This means that natural gas price played a partial mediating role in the GPR's impact on energy exports in emerging economies.

## 5. Heterogeneity Analysis of the Impact of Geopolitics on the Energy Trade

*5.1. The Heterogeneity Analysis Based on the Differences of National Attributes*

5.1.1. Whether an OECD Member or Not

Although the above results can be used to understand the impact of geopolitical risk on the energy trade of emerging economies as a whole, the differences in national attributes were ignored. For sample countries, whether they joined the OECD may have caused differences in the impact of GPR on their energy trade, so it was necessary to analyze the heterogeneity. From the perspective of national attributes, OECD member countries are usually better than non-member countries in an international environment and energy utilization, so the negative effects of geopolitical risks should be smaller. Previous studies usually distinguished developed and developing economies based on whether they were OECD members, and analyzed the heterogeneity on this basis. Sang-Ho and Kim found that developed countries exported more carbon in their trade with OECD countries, but imported more carbon in their trade with non-OECD countries [44]. Developed countries should subsidize underdeveloped countries to develop renewable energy [45]. In recent years, the carbon-intensive behaviors of OECD member countries decreased and they gradually increased the purchase of intermediate products and final products from non-member countries, which is bound to lead to energy trade differences between the two types of countries [46]. Wood et al. analyzed environmental issues in trade and found that OECD member countries had higher energy efficiency than non-member countries [47]. The study by Niu et al. showed that compared with non-member countries, trade openness has a better impact on OECD members [48]. All the above studies indicated that whether an economy was an OECD member had an impact on its energy trade. Therefore, this paper assumed that the geopolitical risk had less negative impact on the energy trade of OECD members. OECD members in the sample of this paper were as follows: South Korea (1996), Mexico (1994), Turkey (1961), Israel (2010), and Colombia (2020). For the time span of this paper, only Israeli (2010 and after) and Colombia (2020 and after) were regarded as OECD members, and the remaining emerging economies had not officially joined the OECD. In this paper, the fixed-effect regression model was adopted to analyze OECD member countries and non-member countries, and the estimated results are shown in Table 10.

Table 10. Test results of the impact of GPR on the energy trade of OECD and non-OECD members.

|  | OECD Members | | Non-OECD Members | |
|---|---|---|---|---|
|  | LnImport | LnExport | LnImport | LnExport |
| LnGPR | 0.065 (1.37) | −0.267 *** (−4.66) | −0.162 *** (−5.55) | −0.114 *** (−3.91) |
| Control Variable | Yes | | Yes | |
| Time Control | Yes | | Yes | |
| Individual Control | Yes | | Yes | |
| N | 852 | 852 | 3157 | 3157 |
| F | 288.82 | 302.12 | 1337.06 | 699.41 |
| R-squared | 0.4196 | 0.6967 | 0.2824 | 0.1367 |

[1] Notes: *, **, *** stand for significant levels of 10%, 5%, and 1%, respectively, and the values in brackets are T-values.

As can be seen from Table 10, the impact of GPR index on the energy trade of emerging economies was indeed heterogeneous in terms of whether they joined the OECD or not. The increase in GPR had no significant negative effect on the energy imports of OECD member countries, and even had a positive effect to some extent, which is consistent with the hypothesis. However, surprisingly, the energy exports of OECD member countries were more strongly affected by the negative impact of geopolitical risks. The reason may be that the energy exports of OECD member countries, except Mexico, were smaller in magnitude and those countries did not have to undertake the task of world energy supply, so their energy exports fluctuated more in the impact of events. The geopolitical risk of non-member countries dampened their energy trade by both imports and exports, consistent with the baseline regression.

5.1.2. Energy Importing Countries and Energy Exporting Countries

Next, we analyzed whether the influence of GPR index on the energy trade of emerging economies was heterogeneous between net energy exporters and net importers. Net energy importing countries have a more passive and inferior position in international energy trade compared to exporting countries, so the negative effects of geopolitical risks should be greater. Hongwei et al., based on a complex network and panel regression analysis, found the country risk had a great influence on the trade pattern of energy importers and exporters [29]. Importing countries should pay attention to the negative impact of economic risk, which will reduce the importer's resource anti-control ability and worsen its relationship with important countries. For exporters, political and economic risks have a negative impact on their resource control ability and the total energy trade volume. If the net export value obtained by subtracting the energy import value from the energy export value of the sample data is positive, the sample is an energy exporting country; otherwise, it is an energy importing country. A fixed-effect regression model was used to analyze the energy importing countries and energy exporting countries, and the estimated results are shown in Table 11.

Table 11. Test results of the impact of GPR on energy importing and exporting countries.

|  | Energy-Importing Countries | | Energy-Exporting Countries | |
|---|---|---|---|---|
|  | LnImport | LnExport | LnImport | LnExport |
| LnGPR | −0.096 *** (−4.13) | −0.261 *** (−5.85) | −0.021 (−0.43) | −0.115 *** (−3.91) |
| Control variable | Yes | | Yes | |
| Time control | Yes | | Yes | |
| Individual control | Yes | | Yes | |
| N | 2280 | 2280 | 1729 | 1729 |
| F | 1449.70 | 526.34 | 565.53 | 593.27 |
| R-squared | 0.5419 | 0.4466 | 0.0472 | 0.3902 |

[1] Notes: \*, \*\*, \*\*\* stand for significant levels of 10%, 5%, and 1%, respectively, and the values in brackets are T-values.

It can be seen from Table 11 that there was heterogeneity in the impact of GPR index on energy-importing countries and energy-exporting countries. The rise of GPR had a significant negative impact on the imports and exports of energy-importing countries with emerging economies. An increase in GPR had an insignificant negative impact on the energy imports of energy-exporting countries with emerging economies and a significant negative impact on their energy exports. In comparison, energy-importing countries were more adversely affected by geopolitical disputes.

*5.2. Heterogeneity of the Impact of Geopolitical Risks on the Energy Trade Based on Different Geopolitical Event Types*

When geopolitical risks occur, emerging economies will make discretionary decisions on energy imports and exports based on the degree of correlation between energy trade and their own interests. As relatively independent decision-making units, different emerging economies have different reactions to the same geopolitical events, and the same emerging economy also has different reactions when facing different geopolitical events. Therefore, the impact of rising geopolitical risk caused by different types of geopolitical events on energy trade of emerging economies is heterogeneous.

The rise of the geopolitical risk index is often closely related to the occurrence of specific geopolitical events. Figure 2 shows the trend of the global GPR index provided by Caldara and Iacoviello after January 2000. It can be found that the GPR index showed great jumps at specific time points, such as 9/11 and the Iraq War. Although the occurrence of different types of events leads to a rise in GPR, the impact of different types of events on energy trade is bound to be heterogeneous, which is analyzed in the following paragraphs.

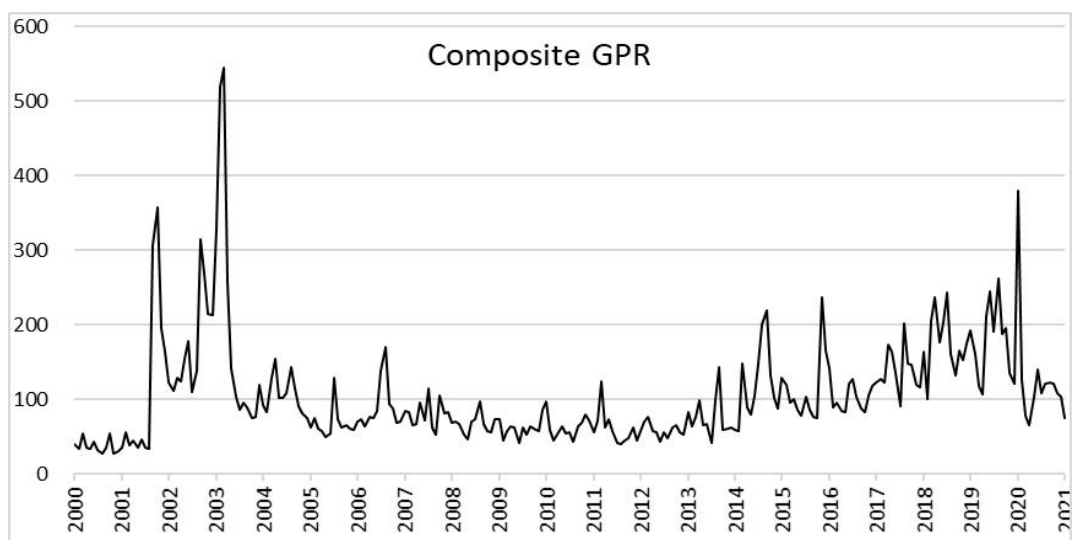

**Figure 2.** Composite GPR trend line.

As can be seen from Figure 2, there was a certain number of "peaks" in the trend line of GPR. In order to consider the heterogeneity of the impact of different types of geo-events, this paper divided events into three categories according to their forms, namely, political events, economic events, and social events.

The RD method was employed to test the impact of different types of geo-events on the energy trade of emerging economies and analyze the heterogeneity. The RDD can effectively analyze the causal relationship between event occurrence and energy imports and exports by using realistic constraint conditions. When the random experiment is not available, the RDD can avoid the endogenous problem of parameter estimation in the analysis of specific geopolitical events so as to truly reflect the causal relationship between the rise of GPR caused by geopolitical events and the energy imports and exports of emerging economies, and the jump effect of events can be used to estimate the causal relationship between them. The accuracy of the RD results is affected by the model setting, to which the bandwidth is the key. In the selection of bandwidth, the IK method was used to calculate the optimal bandwidth (OB) [49]. It is worth noting that the OB in this part was 5–7 months, which can represent the short term; the two times OB was about 10–14 months, which can represent the medium and long term. According to the research of Lee and Lemieux [50], the following model was constructed:

$$\text{LnImport}_{it}/\text{LnExport}_{it} = \lambda_0 + \lambda_1 * \text{Event}_{it} + \lambda_2 * Y_{it} + \lambda * X_{it} + \pi_i + \varepsilon_{it} \qquad (7)$$

$$\text{Event}_{it} = \begin{cases} 1, & Y_{it} > 0 \\ 0, & Y_{it} \leq 0 \end{cases}$$

where i represents an individual economy and t represents time; $\text{Event}_{it}$ is the processing variable, that is, when t is after the occurrence of the event, the value is 1, otherwise it is 0; $Y_{it}$ is the execution variable, that is, the difference between t and the time of the event; X is the covariable, including GDP, interest rate, and exchange rate; $\pi_i$ represents the individual fixed effect; $\varepsilon_{it}$ is the error term; and $\lambda_1$ represents the impact of events on energy imports and exports, which is the main concern coefficient. The RD in this part adopted this model, which will not be repeated. Sensitivity analysis can be used to further enrich and move variable events forward or backward. This can help draw conclusions about the potential anticipated (or postponed) effects of the event [51]. However, because of the length limit, this paper only conducted regression discontinuity for the time point of the event.

In addition, energy prices and elasticity of supply and demand were also taken into account when analyzing the impact of different types of geopolitical events. Combined with the above mediating effect test, it was found that energy prices played a mediating role in the process of geopolitical risks affecting energy trade. The selection of energy prices continued to follow the three energy prices mentioned above, and the monthly trend is shown in Figure 3.

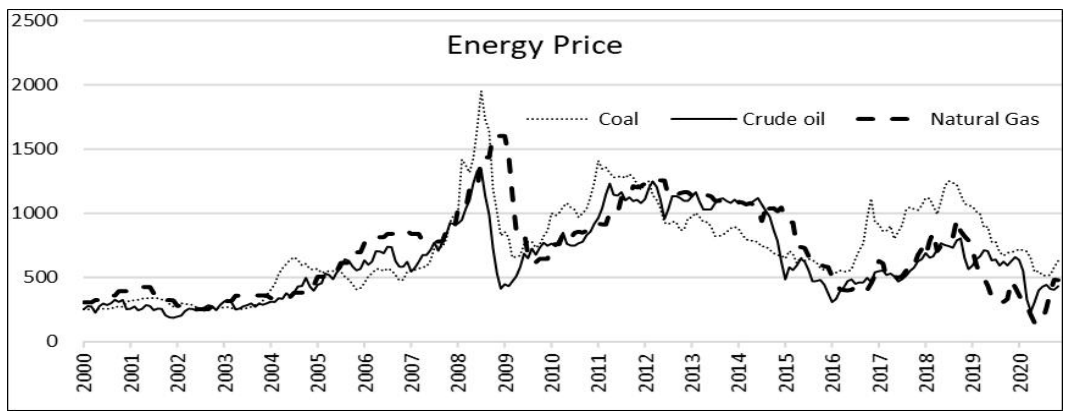

**Figure 3.** Energy price trend line.

Next, this paper carried out the RD for political events, social events, and economic events to analyze the heterogeneity of the impact on energy trade based on the different geo-event types.

5.2.1. RD Analysis of Political Events

This paper selected the Arab Spring as representative of political events and used the RD for analysis. The Arab Spring was a wave of revolutions in the Arab world, with the civil war in Syria and the war in Libya being the main parts. The Syrian civil war usually refers to the conflict between the Syrian government, the Syrian opposition groups, and the Islamic State that started at the beginning of 2011. The anti-government demonstrations in Syria started on 26 January 2011 and escalated on 15 March 2011, and then the anti-government demonstrations evolved into armed conflicts. The Libyan War was an armed conflict that occurred in Libya in 2011, often referred to as the "February 17 Revolution" in Libya. The fighting was between the government, led by Muammar Gaddafi, and the forces against him.

On the energy level, Libya was the most affected by the Arab Spring movement. Its oil production was affected by the twists and turns, which lasted for a year: According to OPEC data, the production of Libyan crude oil was 1.6 million barrels per day (KBD) before January 2011, but it fell sharply after February, with a crude oil output of 375 KBD in March. By July, the crude oil production had reached a bottom, with only 7000 barrels per day. In May 2012, its crude oil production was back to 1441 KBD. On 14 April 2018, the United States, Britain, and France launched air strikes in Syria. The events related to this section were the Syrian and Libyan Wars (March 2011), the escalation of the Syrian War (September 2013), and Syrian tensions (April 2018). This paper conducted a sharp RD analysis of three events, analyzing the heterogeneity of the whole sample, and energy importing countries and energy exporting countries affected by the event before and after the event. Tables 12–14 report the results of the RD, and Figure 4–6 show the energy price changes before and after the event.

**Table 12.** The impact of the Arab Spring.

| Sample | LnImport | | LnExport | |
|---|---|---|---|---|
| | OB | 2 * OB | OB | 2 * OB |
| All | 0.010 | 0.119 * | 0.048 | 0.024 |
| Energy-importing | 0.130 * | 0.160 *** | - | - |
| Energy-exporting | - | - | 0.002 | 0.021 |
| Individual control | Yes | | Yes | |
| Control variable | Yes | | Yes | |

[1] Notes: *, **, *** stand for significant levels of 10%, 5%, and 1%, respectively.

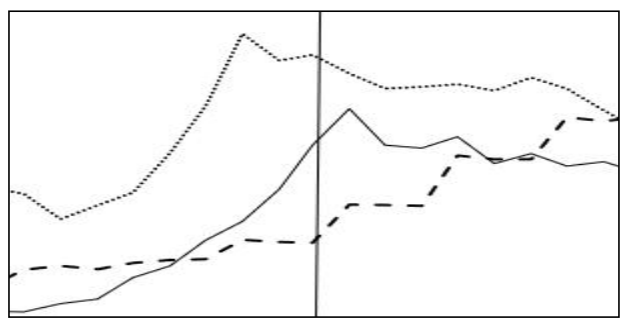

**Figure 4.** Energy price around March 2011.

As can be seen from Table 12, the outbreak of wars in Syria and Libya increased the energy imports of energy-importing countries in the short term, as well as increased the energy imports of all of the sample countries and the energy-importing countries in the long term, and it also promoted energy exports insignificantly. Combined with Figure 4, we found that world energy prices rose sharply throughout the Arab Spring, with crude oil prices peaking in May and coal prices continuing to climb. The rise in energy price in the short term was the direct cause of the increase in energy imports from energy-importing countries in the short term. In the medium and long term, energy prices continued to rise during the turmoil in the Middle East and maintained for a long time after reaching a peak in May, which directly led to a significant increase in energy trade volume in the medium and long term. Energy demand showed a strong rigidity.

**Table 13.** The impact of the Syrian War escalation.

| Sample | LnImport | | LnExport | |
|---|---|---|---|---|
| | OB | 2 * OB | OB | 2 * OB |
| All | 0.026 | −0.006 | −0.036 | −0.019 |
| Energy-importing | 0.013 | −0.055 | - | - |
| Energy-exporting | - | - | 0.009 | 0.033 |
| Individual control | Yes | | Yes | |
| Control variable | Yes | | Yes | |

[1] Notes: *, **, *** stand for significant levels of 10%, 5%, and 1%, respectively.

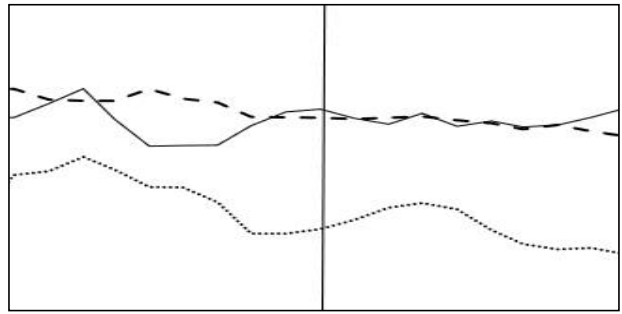

**Figure 5.** Energy price around September 2013.

It can be seen from Table 13 that the escalation of the Syrian War had no significant impact on the energy trade of emerging economies in the short and medium term. The turmoil in the Middle East has lasted for a long time, and neighboring countries have gradually adapted to the political environment. At the same time, Syria is located in the Middle East, but it is not an oil producer, and its escalation has not affected the energy supply in the region. Energy price volatility also flattened, as shown in Figure 5.

**Table 14.** The impact of Syrian tensions.

| Sample | LnImport | | LnExport | |
|---|---|---|---|---|
| | OB | 2 * OB | OB | 2 * OB |
| All | 0.115 ** | 0.108 *** | 0.059 | 0.025 |
| Energy-importing | 0.134 ** | 0.112 ** | - | - |
| Energy-exporting | - | - | 0.010 | 0.051 |
| Individual control | Yes | | Yes | |
| Control variable | Yes | | Yes | |

[1] Notes: *, **, *** stand for significant levels of 10%, 5%, and 1%, respectively.

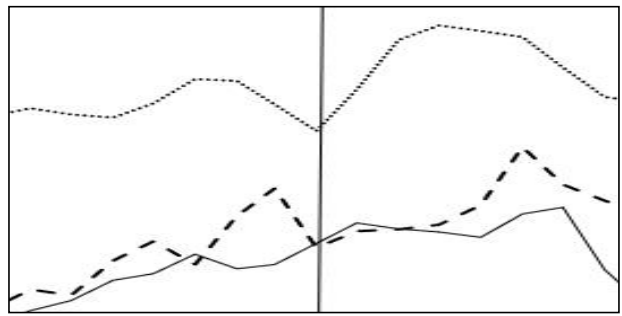

**Figure 6.** Energy price around April 2018.

As can be seen from Table 14, the tension in Syria in 2018 significantly increased energy imports for all of the sample countries and energy-importing countries in the short and medium term, and it also increased energy exports to a certain extent. Figure 6 shows that energy prices continued to rise over the medium to long term, reaching a peak in October 2018. The price rise directly led to the short-term and long-term increase in energy imports. Although the price fell after October, it did not change the jump trend of energy imports before or after the discontinuity.

This paper also analyzed political events such as the Iraq War, the increase in troops in Afghanistan, and Russia's annexation of Crimea, and the conclusion is consistent with the previous analysis. Generally speaking, the occurrence of serious political events tends to lead to regional political instability. Therefore, the price of fossil energy will be increased, thus increasing the energy trade volume of emerging economies.

5.2.2. RD Analysis of Social Events

In this paper, the outbreak of the COVID-19 virus in January 2020 was selected as the representative of social events, and the RD method was used for analysis. The first case of COVID-19 was detected in Wuhan, China, in December 2019. Since then, the disease spread around the world, leading to a continuing pandemic. According to the U.S. Energy Information Administration's (EIA) Annual Energy Outlook 2021 (AEO2021), released on February 3, 2021, it may take several years for the United States to return to 2019 levels of energy consumption and carbon dioxide emissions due to the impact of COVID-19 on the U.S. economy and the global energy sector. EIA Acting Administrator Stephen Nalley said that it will take years for the U.S. energy industry to reach its new normal. In 2020, the epidemic triggered a historic energy demand shock, leading to reductions in greenhouse gas emissions, reduced energy production, and volatile commodity prices. The pace of economic recovery, technological advances, changes in trade flows, and energy incentives will determine how the world produces and consumes energy in the future. In this paper, a sharp RDD was performed to analyze the heterogeneity of the event impacts on the whole sample, as well as energy-importing countries and energy-exporting countries before and after the event. Table 15 reports the results of the regression, and Figure 7 reports the energy price changes before and after the event.

**Table 15.** The impact of COVID-19.

| Sample | LnImport | | LnExport | |
|---|---|---|---|---|
| | OB | 2 * OB | OB | 2 * OB |
| All | −0.169 ** | −0.222 *** | −0.051 | −0.148 ** |
| Energy-importing | −0.241 *** | −0.272 *** | - | - |
| Energy-exporting | - | - | 0.148 | 0.069 |
| Individual control | Yes | | Yes | |
| Control variable | Yes | | Yes | |

[1] Notes: *, **, *** stand for significant levels of 10%, 5%, and 1%, respectively.

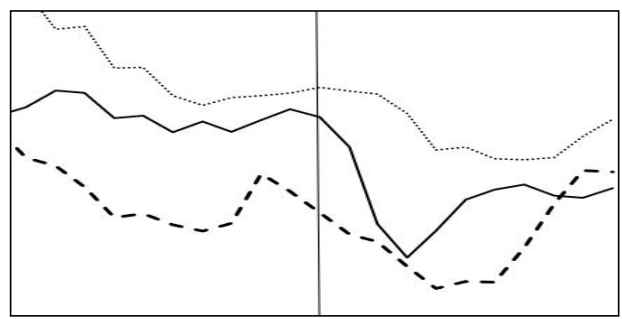

**Figure 7.** Energy price around January 2020.

It can be seen from Table 15 that the impact of the COVID-19 outbreak on energy trade was reflected in the short-, medium-, and long-term inhibition of the overall energy imports of energy-importing countries, while significantly inhibiting the overall energy exports in the medium and long term. At the same time, it can be found from Figure 7 that energy prices fell sharply near the discontinuity, which was the direct reason for the decline in energy imports and exports in the short term. In addition, due to the obstacles of trade circulation caused by the epidemic and the uncertainty of the future of production activities, energy demand was bound to be lower than before the outbreak.

This paper also analyzed social events such as the Fukushima nuclear leak and the Ebola outbreak, and the conclusions are largely the same as the previous ones. In general, when serious social events occur, public safety is seriously threatened. The collapse of energy prices and the sharp decline in demand together lead to a decline in energy trade volume.

5.2.3. RD Analysis of Economic Events

In this paper, the United States–China trade friction was selected as representative of economic events, and the RD method was used for analysis. On 1 August 2019, due to the Trump administration's dissatisfaction with the Chinese government's purchase process of U.S. agricultural products, former President Trump announced on Twitter that he would impose a 10% tariff on all remaining USD 300 billion of Chinese imports to the US starting from 1 September 2019. On 5 August, the RMB exchange rate against the USD fell below 7. On the same day, the US Treasury announced that China was listed as a currency manipulator. Subsequently, the Chinese government announced a suspension of the purchase of American agricultural products. On 24 August, it announced additional tariffs of 10% or 5% on USD 75 billion of U.S. goods and resumed additional tariffs on U.S. cars and parts. The United States responded the next day by imposing a 15% tariff on USD 300 billion of Chinese goods and a 25% tariff to 30% tariff on USD 250 billion of Chinese goods, which was later shelved. In this paper, a sharp RDD was performed to analyze the heterogeneity of the event impacts on the whole sample, as well as energy-importing countries and energy-exporting countries before and after the event. Table 16 reports the results of the regression, and Figure 8 reports the energy price changes before and after the event.

**Table 16.** The impact of U.S.–China tensions.

| Sample | LnImport | | LnExport | |
|---|---|---|---|---|
| | OB | 2 * OB | OB | 2 * OB |
| All | −0.093 * | −0.021 | 0.009 | −0.040 |
| Energy−importing | −0.131 ** | −0.047 | − | − |
| Energy−exporting | − | − | 0.006 | −0.062 |
| Individual control | Yes | | Yes | |
| Control variable | Yes | | Yes | |

[1] Notes: *, **, *** stand for significant levels of 10%, 5%, and 1%, respectively.

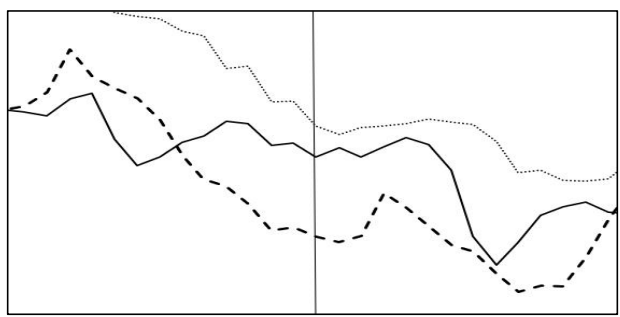

**Figure 8.** Energy prices around August 2019.

As can be seen from Table 16, the impact of the U.S.–China trade friction on energy trade was reflected in the short-term reduction in the total sample and energy imports of energy-importing countries, and in the medium and long term, there was a certain inhibitory effect on energy imports and exports. The World Bank issued a report on 29 October 2019, saying that, affected by the decline in demand caused by the weak global economic growth prospects, the prices of energy, metals, and other commodities would "fall sharply" in 2019, as evidenced by the significant decline in energy prices in Figure 8. John Yergin, Chairman of Cambridge Energy Research Associates, said that at present, the growth in China's demand is the basic driving force to promote the world oil market. The trade frictions between China and the United States also dampened China's energy imports to some extent.

After that, we analyzed the global economic crisis caused by the 2008 financial crisis and the subsequent European debt crisis, and this paper reached the same conclusion as above. In general, the occurrence of serious economic events will generally lead to the suppression of the energy trade, and the decline in prices and trading volume is the main reason for the short-term decline in energy trade volume.

In conclusion, there is indeed heterogeneity in the impact of geopolitical risks caused by specific events on the energy trade of emerging economies, which is reflected in the heterogeneity of the types of geopolitical events. First, political events lead to rising energy trade by increasing energy prices. International political events usually accompany the threat of war between countries or even direct conflicts, bringing serious tension to the region. This has caused the shortage of energy supplies to a large extent, and the price of energy increases with it, but the rigidity of energy demand prevents the imports from decreasing significantly. Second, social events lead to a decline in energy trade volume by reducing energy prices and demand. Major social events lead to a cessation of normal domestic production activities and a reduction in energy demand. Meanwhile, the uncertainty of the future situation also leads to a withdrawal in the long-term in the spot and futures market and falling prices. Finally, economic events lead to a decrease in energy trade volume by reducing energy prices and energy transaction volume. Modern major economic events include international trade friction between the big powers or international trade sanctions, along with the limitations of commodity imports and exports and the deterioration of business environment, with which the energy trade decreases significantly and energy prices keep falling.

## 6. Conclusions and Implications

Firstly, rising geopolitical risks have a significant negative impact on the energy trade of emerging economies. On the one hand, rising geopolitical risks have significantly reduced energy imports and exports in emerging economies. On the other hand, the rising geopolitical risk has a more negative effect on the energy exports of emerging economies than on the energy imports. Besides, the inhibitory effect still exists after excluding non-political factors, sample size, and grouping bias.

Secondly, there is a time-lag effect and a mediating effect on the impact of the rising geopolitical risks on emerging economies' energy trade. First, the negative impact on energy exports and imports shows a short-term and medium-term lag effect, in which the lag effect of imports is weak in the short term, and the short-term and medium-term impact shows an upward trend, and the impact on imports is smaller than that on exports. Second, the inhibition of geopolitical risks on the energy imports and exports of emerging economies is partly realized by reducing the prices of coal and crude oil and partly achieved by raising the price of natural gas.

Finally, the impact of geopolitical risks on energy trade is heterogeneous, which is reflected in the differences of national attributes and types of geopolitical events. First, compared to non-OECD member countries, the energy imports of OECD member countries had no significant positive performance when GPR increased, but the energy exports decreased more; second, the net energy importing countries were more negatively affected when the geopolitical risks rose; third, political events usually led to an increase in energy trade, whereas social events and economic events usually led to a decrease in energy trade.

The conclusions of this paper can be used as a reference for the energy trade strategy of emerging economies. First of all, the analysis of this paper shows that geopolitical risk has a negative impact on the energy trade of emerging economies from all perspectives. Therefore, emerging economies should try to avoid geopolitical risks in the international energy trade and work together to create a stable and harmonious international environment. Second, energy-importing and -exporting countries should make targeted strategic adjustments according to their own situations. The empirical results show that importing countries are more negatively affected by geopolitical risks than the exporting countries. They should enrich their own import channels, reduce the dependence on energy imports from specific countries, strengthen their own anti-control abilities of resources, and consider relying on international cooperation organizations to enhance their status in the energy market if conditions permit. Energy exporters also suffer losses in geopolitical risks, which can be explained by the inhibition of energy exports of emerging economies faced with geopolitical risks. Therefore, energy-exporting countries should enrich their export channels, combining fiscal policies to expand domestic demand to reduce their dependence on foreign trade, stimulating domestic consumption, driving industrial growth, and promoting internal energy consumption. Finally, emerging economies should formulate appropriate policy strategies in the face of different types of geopolitical events [52], and make discretionary decisions in terms of energy import and export in the long term and short term. Overall, emerging economies need to focus on the following three perspectives: implementing short-term, medium-term, or long-term regulatory policies and strategies on the trade, diplomacy, and supply side; improving the pertinence of response measures and gradually stabilizing their cooperation in the international energy market; and maintaining the harmony of geopolitical relations in a long-term and stable way.

**Author Contributions:** Conceptualization, Z.L.; data curation, C.Y.; formal analysis, C.Y.; funding acquisition, F.L.; methodology, Z.L.; project administration, F.L. and P.F.; resources, C.Y.; software, C.Y.; supervision, F.L. and Z.L.; validation, P.F.; writing—original draft, C.Y.; writing—review and editing, C.Y., Z.L., and P.F. All authors have read and agreed to the published version of the manuscript.

**Funding:** This research was funded by Teaching Reform Research Project of Higher Education in Hunan Province "Exploration on the 'Trinity' Mixed Teaching Mode Reform of Ideological and Political Theory Course in Local Universities from the Perspecjutive of 'Classroom Revolution'" (Xiangjiaotong [2019] No. 291)

**Institutional Review Board Statement:** Not applicable.

**Informed Consent Statement:** Not applicable.

**Data Availability Statement:** Not applicable.

**Acknowledgments:** The authors would like to thank the anonymous reviewers who provided many helpful suggestions for improvements.

**Conflicts of Interest:** The authors declare no conflict of interest.

## References

1. Yang, Y.; Li, J.; Sun, X.; Chen, J., Measuring External Oil Supply Risk: A Modified Diversification Index with Country Risk and Potential Oil Exports. *Energy* **2014**, *68*, 930−938.
2. Lee, C.-C.; Lee, C.-C.; Ning, S.-L. Dynamic Relationship of Oil Price Shocks and Country Risks. *Energy Econ.* **2017**, *66*, 571−581.
3. Duan, F.; Ji, Q.; Liu, B.-Y.; Fan, Y. Energy Investment Risk Assessment for Nations along China's Belt & Road Initiative. *J. Clean. Prod.* **2018**, *170*, 535−547.
4. Kang, W.; Ratti, R.A. Structural Oil Price Shocks and Policy Uncertainty. *Econ. Model.* **2013**, *35*, 314−319.
5. Antonakakis, N.; Chatziantoniou, I.; Filis, G. Dynamic Spillovers of Oil Price Shocks and Economic Policy Uncertainty. *Energy Econ.* **2014**, *44*, 433−447.
6. Sun, M.; Gao, C.; Shen, B. Quantifying China's Oil Import Risks and the Impact on the National Economy. *Energy Policy* **2014**, *67*, 605−611.
7. Zhang, H.-Y.; Xi, W.-W.; Ji, Q.; Zhang, Q. Exploring the Driving Factors of Global LNG Trade Flows Using Gravity Modelling. *J. Clean. Prod.* **2018**, *172*, 508−515.
8. Gupta, E. Oil Vulnerability Index of Oil−importing Countries. *Energy Policy* **2008**, *36*, 1195−1211.

9. Le Coq, C.; Paltseva, E. Measuring the Security of External Energy Supply in the European Union. *Energy Policy* **2009**, *37*, 4474–4481.
10. Liu, Y.; Li, Z.; Xu, M. The Influential Factors of Financial Cycle Spillover: Evidence from China. *Emerg. Mark. Financ. Trade* **2020**, *56*, 1336–1350.
11. Pena, G. A New Trading Algorithm with Financial Applications. *Quant. Financ. Econ.* **2020**, *4*, 596–607.
12. Rui Xie, Y.Z. Liming Chen, Structural Path Analysis and Its Applications: Literature Review. *Natl. Account. Rev.* **2020**, *2*, 83–94.
13. Mercille, J.; Jones, A. Practicing Radical Geopolitics: Logics of Power and the Iranian Nuclear oCrisiso. *Ann. Assoc. Am. Geogr.* **2009**, *99*, 856–862.
14. Checkovich, A. American Empire: Roosevelt's Geographer and the Prelude to Globalization. *Isis* **2005**, *96*, 455–456.
15. Harvey, T. The Limits to Capital, 3rd Edition. *Prog. Hum. Geogr.* **2008**, *32*, 485–486.
16. Kurecic, P. The Three Seas Initiative: Geographical Determinants, Geopolitical Foundations, and Prospective Challenges. *Croat. Geogr. Bull.* **2018**, *80*, 99–124.
17. Serbos, S.; Anastasiadis, G. Revisiting Europe's Geopolitical Landscape after the Ukraine Crisis: America's Balance of Power Strategy. *Rev. UNISCI* **2018**, *46*, 177–195.
18. Kemfert, C.; Schmalz, S. Sustainable Finance: Political Challenges of Development and Implementation of Framework Conditions. *Green Financ.* **2019**, *1*, 237–248.
19. Pant, B.; Rai, R.K.; Bhattarai, S.; Neupane, N.; Kotru, R.; Pyakurel, D. Actors in Customary and Modern Trade of Caterpillar Fungus in Nepalese High Mountains: Who Holds the Power? *Green Financ.* **2020**, *2*, 373–391.
20. Ferdinand, P. Westward Ho−the China Dream and 'One Belt, One Road': Chinese Foreign Policy under Xi Jinping. *Int. Aff.* **2016**, *92*, 941–957.
21. Marechal, N. Networked Authoritarianism and the Geopolitics of Information: Understanding Russian Internet Policy. *Media Commun.* **2017**, *5*, 29–41.
22. Oral, M.; Ozdemir, U. The Position of Turkey in Global Energy Geopolitics: Opportunities and Risks. *J. Hist. Cult. Art Res.* **2017**, *6*, 948–959.
23. Zhao, G. China−U.S.–India: Is a New Triangle Taking Shape? *China Q. Int. Strateg. Stud.* **2016**, *2*, 1–16.
24. Suk, H.W. International Political Implications of Ukraine Crisis 2014 and the Korean Peninsula. *Korean J. Slav. Stud.* **2014**, *30*, 89–118.
25. Simonia, N.A.; Torkunov, A.V. The Impact of Geopolitical Factors on International Energy Markets (the US Case). *Polis. Political Stud.* **2016**, *2*, 38–48.
26. Oswald, U. Energy Security, Availability, and Sustainability in Mexico. *Rev. Mex. Cienc. Politicas Soc.* **2017**, *62*, 155–195.
27. dos Santos, M.M.; Lara dos Santos Matai, P.H. The Importance of the Industrialization of Brazilian Shale When Faced with the World Energy Scenario. *Rev. Esc. Minas* **2010**, *63*, 673–678.
28. Zhang, H.−Y.; Ji, Q.; Fan, Y. What Drives the Formation of Global Oil Trade Patterns? *Energy Econ.* **2015**, *49*, 639–648.
29. Hongwei, Z.; Ying, W.; Cai, Y.; Yaoqi, G. The Impact of Country Risk on Energy Trade Patterns Based on Complex Network and Panel Regression Analyses. *Energy* **2021**, *222*, 119979.
30. Santeramo, F.G.; Cioffi, A. The Entry Price Threshold in EU Agriculture: Deterrent or Barrier? *J. Policy Modeling* **2012**, *34*, 691–704.
31. Dal Bianco, A.; Boatto, V.L.; Caracciolo, F.; Santeramo, F.G. Tariffs and Non−tariff Frictions in the World Wine Trade. *Eur. Rev. Agric. Econ.* **2016**, *43*, 31–57.
32. Beghin, J.C.; Schweizer, H. Agricultural Trade CostsJEL Codes. *Appl. Econ. Perspect. Policy* **2020**, doi:10.1002/aepp.13124.
33. Caldara, D.; Iacoviello, M. Measuring Geopolitical Risk. In *Board of Governors of the Federal Reserve Board*; Federal Reserve: Washington, DC, USA, 2019.
34. Zawadzki, K. The Performance of ETFs on Developed and Emerging Markets with Consideration of Regional Diversity. *Quant. Financ. Econ.* **2020**, *4*, 515–525.
35. An, H.; Gao, X.; Fang, W.; Huang, X.; Ding, Y. The Role of Fluctuating Modes of Autocorrelation in Crude Oil Prices. *Phys. A Stat. Mech. Appl.* **2014**, *393*, 382–390.
36. Zhong, W.; An, H.; Fang, W.; Gao, X.; Dong, D. Features and Evolution of International Fossil Fuel Trade Network Based on Value of Emergy. *Appl. Energy* **2016**, *165*, 868–877.
37. Alhajeri, N.S.; Dannoun, M.; Alrashed, A.; Aly, A.Z. Environmental and Economic Impacts of Increased Utilization of Natural Gas in the Electric Power Generation Sector: Evaluating the Benefits and Trade−offs of Fuel Switching. *J. Nat. Gas Sci. Eng.* **2019**, *71*, 102969.
38. Li, Z.; Liao, G.; Albitar, K. Does Corporate Environmental Responsibility Engagement Affect Firm Value? The Mediating Role of Corporate Innovation. *Bus. Strategy Environ.* **2020**, *29*, 1045–1055.

39. Lin, B.-Q.; Liu, J.-H. Estimating Coal Production Peak and Trends of Coal Imports in China. *Energy Policy* **2010**, *38*, 512–519.
40. Li, F.; Huang, Z.; Zhong, J.; Albitar, K. Do Tense Geopolitical Factors Drive Crude Oil Prices? *Energies* **2020**, *13*, 4277.
41. Faria, J.R.; Mollick, A.V.; Albuquerque, P.H.; Leon−Ledesma, M.A. The Effect of Oil Price on China's Exports. *China Econ. Rev.* **2009**, *20*, 793−805.
42. Labandeira, X.; Labeaga, J.M.; Lopez−Otero, X. A Meta−analysis on the Price Elasticity of Energy Demand. *Energy Policy* **2017**, *102*, 549−568.
43. Dagher, L. Natural Gas Demand at the Utility Level: An Application of Dynamic Elasticities. *Energy Econ.* **2012**, *34*, 961−969.
44. Sang-Ho, L.; Kim, C. The Analysis on the Embodied Carbon Trade Using Multi−Regional Input−Output Model. *Korea Trade Rev.* **2012**, *37*, 1−19.
45. Ilic, B.; Stojanovic, D.; Djukic, G. Green Economy: Mobilization of International Capital for Financing Projects of Renewable Energy Sources. *Green Financ.* **2019**, *1*, 94−109.
46. Jiang, X.M.; Guan, D.B. The Global $CO_2$ Emissions Growth after International Crisis and the Role of International Trade. *Energy Policy* **2017**, *109*, 734−746.
47. Wood, R.; Stadler, K.; Simas, M.; Bulavskaya, T.; Giljum, S.; Lutter, S.; Tukker, A. Growth in Environmental Footprints and Environmental Impacts Embodied in Trade: Resource Efficiency Indicators from EXIOBASE3. *J. Ind. Ecol.* **2018**, *22*, 553−564.
48. Niu, J.; Wen, J.; Yang, X.-Y.; Chang, C.-P. Trade Openness, Political Stability and Environmental Performance: What Kind of Long−Run Relationship? *Probl. Ekorozw.* **2018**, *13*, 57−66.
49. Imbens, G.; Kalyanaraman, K. Optimal Bandwidth Choice for the Regression Discontinuity Estimator. *Rev. Econ. Stud.* **2012**, *79*, 933−959.
50. Lee, D.S.; Lemieux, T. Regression Discontinuity Designs in Economics. *J. Econ. Lit.* **2010**, *48*, 281−355.
51. Jayasinghe, S.; Beghin, J.C.; Moschini, G. Determinants of World Demand for US Corn Seeds: The Role of Trade Costs. *Am. J. Agric. Econ.* **2010**, *92*, 999−1010.
52. Li, T.; Zhong, J.; Huang, Z. Potential Dependence of Financial Cycles between Emerging and Developed Countries: Based on ARIMA-GARCH Copula Model. *Emerg. Mark. Financ. Trade* **2020,** *56*, 1237-1250.